\documentclass[aps,prd,amssymb,amsfonts,twocolumn,showpacs,preprintnumbers]{revtex4}
\usepackage{amsmath,amssymb}
\usepackage[dvips]{graphicx}

\newcommand{\beeq}{\begin{equation}}
\newcommand{\eneq}{\end{equation}}
\newcommand{\beeqa}{\begin{eqnarray}}
\newcommand{\eneqa}{\end{eqnarray}}
\newcommand{\cH}{\mathcal{H}}
\newcommand{\cR}{{\cal R}}
\newcommand{\dd}{\partial}

\begin{document}


\title
{Primordial magnetic fields from second-order cosmological perturbations:\\ Tight coupling approximation}

\author{Satoshi~Maeda$^{1,2}$, Satoshi~Kitagawa$^3$, Tsutomu~Kobayashi$^3$ and Tetsuya~Shiromizu$^{1,2}$}
\address{$^1$ Department of Physics, Kyoto University, Kyoto 606-8502, Japan}
\address{$^2$ Department of Physics, Tokyo Institute of Technology, Tokyo 152-8551, Japan}
\address{$^3$ Department of Physics, Waseda University, Tokyo 169-8555, Japan}

\email{smaeda"at"tap.scphys.kyoto-u.ac.jp;
satoshi"at"gravity.phys.waseda.ac.jp;
tsutomu"at"gravity.phys.waseda.ac.jp; shiromizu"at"tap.scphys.kyoto-u.ac.jp}

\begin{abstract}
We explore the possibility of generating large-scale magnetic fields from second-order
cosmological perturbations during the pre-recombination era.
The key process for this is Thomson scattering
between the photons and the charged particles within the cosmic plasma.
To tame the multi-component interacting fluid system,
we employ the tight coupling approximation.
It is shown that
the source term for the magnetic field is given by
the vorticity, which signals the intrinsically second-order quantities, and
the product of the first order perturbations.
The vorticity itself is sourced by the product of the first-order quantities in the vorticity evolution equation.
The magnetic fields generated by this process are estimated to be $\sim 10^{-29}\,$Gauss
on the horizon scale at the recombination epoch. Although our rough estimate suggests that the current 
generation mechanism can work even on smaller scales, more careful investigation is needed to make
clear whether it indeed works in a wide range of spatial scales. 
\end{abstract}

\preprint{WU-AP/287/08}


\maketitle

\section{Introduction}

Magnetic fields are known to be present on various scales in the universe \cite{Widrow:2002ud}. 
For example, 
magnetic fields are observed in galaxies and clusters, with intensity $\sim 1\;$$\mu$Gauss.
Only an upper limit has been given for magnetic fields on cosmological
scales, $<10^{-9}\;$Gauss.
Primordial large-scale magnetic fields may be present and
serve as seeds for the magnetic fields in galaxies and 
clusters, which are amplified through the dynamo mechanism
after galaxy formation.
We need numerical calculations in order to evaluate precisely
a necessary seed magnetic field, however we can estimate order of
magnitude of it and it is $\sim 10^{-21} - 10^{-31}$ Gauss at $100$Mpc
~\cite{Davis:1999bt}. 
If sufficiently large, the primordial magnetic field
could leave observable imprints on the cosmic microwave background~\cite{Yamazaki:2008gr}.

A number of models 
have been proposed for generating large-scale magnetic fields in the early
universe~\cite{Grosso, defects, Bamba}. 
However, they rely more or less on some unknown physics. 
In the present paper, we discuss magnetogenesis in the 
pre-recombination era using only the conventional physics that has been established. 
The generation of magnetic fields in this era has been studied in
Refs.~\cite{magre, Maartens:2007, Matarrese, Siegel, Kobayashi:2007wd, Ichiki1, Ichiki2, Takahashi:2007ds, Takahashi:2008gn}.
Now it is widely accepted that
large-scale cosmological perturbations, generated from inflation 
in the early universe,
evolve into a variety of structures such as the cosmic microwave background anisotropies and galaxies. 
However, inflation produces only density fluctuations (scalar perturbations) and
gravitational waves (tensor perturbations); vector perturbations, and hence large-scale magnetic fields,
are not generated in the context of usual inflationary scenarios,
unless, for instance, modifying Maxwell theory of electromagnetism. 
Even if they were generated due to some mechanism, they only decays
without any sources.
This no-go argument is based on {\em linear} perturbation theory,
and so we will be studying second-order perturbations
to overcome this difficulty within inflation and standard Maxwell theory.
We consider a multi-fluid system composed
of photons, electrons, and protons,
which are
tightly coupled via Thomson and Coulomb scattering 
but slightly deviate from each other~\cite{Peebles:1970ag}.
Thomson scattering is important for the generation of large-scale magnetic fields
in the pre-recombination era,
because a rotational current will be produced
by the momentum transfer due to the
Thomson interaction.
We shall see how this process occurs by doing the tight coupling expansion.

The time scales of Thomson scattering, Coulomb scattering, and the plasma oscillation
at $z \sim 10^5$ are 
$\tau_T = 10^{3}\;{\rm sec}$, $\tau_C = 10^{-3}\;{\rm sec}$,
and
$\omega_p^{-1} = 10^{-9}\;{\rm sec}$, respectively \cite{Takahashi:2007ds}. 
The last two time scales are much smaller than 
the Thomson time scale, implying that the effects of Coulomb scattering and plasma 
oscillations are smoothed out within the Thomson time scale.
For this reason, 
we will not take seriously Coulomb scattering and plasma oscillations \cite{Kobayashi:2007wd,Takahashi:2007ds}.
And the effect of the synthesis of hydrogen, i.e., the ionization rate, on magnetogenesis may be
taken into account carefully~\cite{Takahashi:2008gn}. 
It has been reported that the change of the ionization rate does not affect the
generation of magnetic fields on cosmological scales, though
the effect can no longer be ignored on scales smaller than AU.
Thus, for our 
current purpose, it is safe to ignore the effect of the synthesis of hydrogen.

This paper is most closely related to the work of Ref.~\cite{Kobayashi:2007wd},
which studied how magnetic fields are generated from nonlinear cosmic inhomogeneities using the
tight coupling approximation.
While the covariant formalism was used in Ref.~\cite{Kobayashi:2007wd},
we use in this paper cosmological perturbation theory and so the two works
are complementary to each other.
Recently, Takahashi {\em et al.} took a simplified Newtonian approach
to see the detailed physical processes that result in magnetogenesis
in the pre-recombination era~\cite{Takahashi:2007ds}.
The present study may be viewed as the general relativistic extension of Ref.~\cite{Takahashi:2007ds}.
Second-order vector perturbations have also been studied previously
in different contexts~\cite{Mena:2007ve, Lu:2007cj, Dunsby, cp_j}. Although our rough estimate 
suggests that the current generation mechanism can work, more careful investigation is needed to make 
clear whether it indeed works in a wide range of spatial scales.

The organization of this paper is as follows. 
In the next section we define our key quantities and derive 
the equations of motion based on
second-order cosmological perturbation theory. 
In Sec.~3, we invoke the tight coupling approximation to 
solve the relevant equations, giving a compact formula describing the generation of magnetic fields. 
In Sec.~4 we give a summary of the present paper.  

\section{Basic equations in second-order cosmological perturbation theory}

\subsection{The energy-momentum equations up to second-order in perturbation theory}


We start with defining the metric and the energy-momentum tensor we use.
Throughout the paper we rely on the energy-momentum (non)conservation equations
and so we explicitly write them down in this section.

The background spacetime is given by the spatially flat
Friedmann-Lemaitre-Robertson-Walker metric.
We write the perturbed metric in the Poisson gauge as
\begin{eqnarray}
ds^2 = a^2(\eta )\left[-\left(1+2\phi\right)d\eta^2 
       + 2\chi_id\eta dx^i\right.
\nonumber\\ \left. 
+  \left( 1-2\cR \right)\delta _{ij} dx^idx^j\right] ,
\label{metric}
\end{eqnarray}
where $a$ is the scale factor and $\eta$ is the conformal time.
We study the cosmological perturbations up to second order,
and hence we write
\begin{eqnarray}
\phi = \phi^{(1)}+\phi^{(2)}, ~
\chi_i = \chi_i^{(2)}, ~
\cR = \cR^{(1)}+\cR^{(2)},
\end{eqnarray}
where we have dropped the first order vector perturbations $\chi_i^{(1)}$
since they are not generated from inflation in the standard scenarios.
We also neglect the tensor perturbations (gravitational waves) for
simplicity.

We consider a multi-fluid system composed of photons $(\gamma)$, electrons $(e)$, and protons $(p)$.
We assume that the energy-momentum tensor for each fluid component is 
given by that of a perfect fluid (i.e., we neglect anisotropic stresses):
\begin{equation}
T_{(I)\mu}^{\;\;\;\nu} = (\rho_I +p_I)u_{(I)\mu} u_{(I)}^\nu + p_I\delta_\mu^{~\nu} 
\qquad (I=\gamma, p, e),
\end{equation}
where $p_I=w_I \rho_I$ ($w_p=w_e=0, w_\gamma=1/3$) and $u^{\mu}_{(I)}$ is the 4-velocity of the fluid
satisfying $g^{\mu\nu}u_{(I)\mu}u_{(I)\nu}=-1$:
\begin{eqnarray}
 u_{(I)}
 ^{\mu} &=& \frac{1}{a}\left(1-\phi+\frac{3}{2}\phi^2+\frac{1}{2}v_{(I)j}v_{(I)}^j , v_{(I)}^i \right),  \label{four1}\\
 u_{(I)\mu} &=& 
     a\Bigl(-1-\phi+\frac{1}{2}\phi^2-\frac{1}{2}v_{(I)j}v_{(I)}^j, 
\Bigr.
\nonumber\\\Bigl.
 & &\qquad\qquad\qquad v_{(I)i}+\chi_i-2\cR v_{(I)i} \Bigr).
\label{four2}
\end{eqnarray}
The energy-momentum tensor is explicitly written as 
\begin{eqnarray}
 T_{(I)0}{}^{0} &=& -\rho_I-\left(\rho_I+p_I\right)v_{(I)j}v_{(I)}^j , \\
 T_{(I)0}{}^{i} &=& -\left(\rho_I+p_I\right)(1+\phi)v_{(I)}^i , \\
 T_{(I)i}{}^{0} &=& \left(\rho_I+p_I\right)\left(v_{(I)i}+\chi_i\right)
\nonumber\\& &\qquad\qquad
-\left(\rho_I+p_I\right)\left(\phi+2\cR\right)v_{(I)i},
\end{eqnarray}
and
\begin{equation}
T_{(I)i}{}^{j} = \left(\rho_I+p_I\right)v_{(I)i}v_{(I)}^j + p_I\delta_i^{~j}.
\end{equation}
Note here that the energy density and pressure of each fluid,
$\rho_I$ and $p_I$, already include inhomogeneous parts: $\rho_I=\rho_I(t, \mathbf{x})$
and $p_I=p_I(t, \mathbf{x})$.
We will do the perturbative expansion of these quantities later.

The energy-momentum tensor of the electromagnetic field is
written in terms of the field strength $F_{\mu\nu}$ as 
\begin{equation}
T_{{\rm EM}}^{\mu\nu} = F^{\mu\lambda }F^{\nu}{}_{\lambda}
                  - \frac{1}{4}g^{\mu\nu}F_{\lambda \sigma}F^{\lambda \sigma}.
\end{equation}
As usual, the field strength is decomposed into 
the electric and magnetic fields as 
\begin{eqnarray}
E^\mu = F^{\mu\nu}u_{(\gamma) \nu} , \quad
B^\mu = \frac{1}{2}\varepsilon ^{\mu\nu\lambda}F_{\nu\lambda},
\end{eqnarray}
where $\varepsilon ^{\mu\nu\lambda}=\eta^{\mu\nu\lambda\sigma}u_{(\gamma)\sigma}$ with 
$\eta^{0123}=1/\sqrt{-g}$. Here we tacitly adopt the photon frame. 

The vorticity, $\omega_{I}^i\equiv-\frac{1}{2}\varepsilon^{i\nu\rho\lambda}u_{(I)\lambda}\nabla_\nu u_{(I)\rho}$, is the important quantity in our discussion.
The vorticity is given explicitly by~\cite{Matarrese}
\begin{eqnarray}
\omega_I^{(1)i} &=&  -\frac{1}{2a^2}\epsilon^{ijk}\dd_jv_{(I)k}^{(1)},
\\
\omega_I^{(2)i} &=& -\frac{1}{2a^2}\epsilon^{ijk}
                      \left[\dd_j\left(v_{(I)k}^{(2)} + \chi_{k}^{(2)}\right)
\right.
\nonumber\\& &\left.
    + v_{(I)j}^{(1)}v_{(I)k}^{(1)'} +v_{(I)j}^{(1)}\dd_k\left(\phi^{(1)}+2\cR^{(1)}\right)\right],
\end{eqnarray}
where $\epsilon_{ijk}$ is the 3-dimensional flat alternating tensor.
Since we are assuming that the first order vector perturbations vanish, we have $\omega^{(1)i}_I=0$.

During the pre-recombination era photons and charged particles are 
tightly coupled via Thomson scattering, and charged particles are tightly coupled
to each other via Coulomb scattering. 
The momentum transfer between fluids $I$ and $J$ are specified by $\kappa^{IJ}_i$.
The momentum transfer due to Thomson scattering is given by
\begin{eqnarray}
\kappa^{\gamma e}_i &= -\kappa^{e\gamma}_i 
= -\frac{4}{3}\sigma_T n_e\rho_\gamma \left(u_{(\gamma) i} -u_{(e)i}\right) ,
\label{tr1}
\\
\kappa^{\gamma p}_i &= -\kappa^{p\gamma}_i = 
-\frac{m_e^2}{m_p^2}\frac{4}{3}\sigma_T n_p\rho_\gamma \left(u_{(\gamma) i} -u_{(p)i}\right) ,
\label{tr2}
\end{eqnarray}
where $\sigma_T$, $\rho_{\gamma}$, $n_e$, $n_p$ and $m_e$, $m_p$ are 
the Thomson cross section, the energy density of photons, the 
number density and mass of electrons and protons \cite{Kobayashi:2007wd}.
Each number density is defined in the rest frame.
The difference between frames in the momentum transfer terms
appears at third order (in the form of the Lorentz factor), and hence can be neglected.
The momentum transfer rate follows from the collision term in the Boltzmann
equation.
A careful and detailed derivation of Eqs.~(\ref{tr1}) and~(\ref{tr2})
(valid at second order) can be found in
Refs.~\cite{Takahashi:2007ds,Pitrou:2007jy, Bartolo:2006cu}. 
The momentum transfer due to Coulomb scattering is written as
\begin{equation}
\kappa^{pe}_i = -\kappa^{ep}_i = -e^2n_pn_e\eta_C \left(u_{(p) i} -u_{(e)i}\right) ,
\end{equation}
where $\eta_C$ is a electric resistivity which comes from Coulomb
scattering \cite{Takahashi:2007ds}.
To derive the energy conservation law and equations of motion, we
compute the divergence of the energy momentum tensor:
\begin{widetext}
\begin{eqnarray}
\nabla_\nu T_{(I)0}{}^{\nu}
&=&  -\left[\rho_I' +3\cH(\rho_I +p_I)\right]
     +3(\rho_I+p_I)\cR' 
-(\rho_I+p_I)\dd_i v_{(I)}^i, \label{En1}
\\
\nabla_\nu T_{(I)i}{}^{\nu}
&=&   \left[(\rho_I+p_I)(v_{(I)i}+\chi_i)\right]' +4\cH(\rho_I+p_I)(v_{(I)i}+\chi_i)
+(\rho_I+p_I)\dd_i(\phi-\phi^2) +\dd_i p_I
\nonumber \\&&
 -\left(\phi+2\cR\right)\left\{ \left[\left(\rho_I+p_I\right) v_{(I)i}\right]'
 +4\cH\left(\rho_I+p_I\right) v_{(I)i}\right\}
 -5(\rho_I+p_I)\cR'v_{(I)i} +(\rho_I+p_I)\dd_j\left(v_{(I)i}v_{(I)}^j\right),
\end{eqnarray}
\end{widetext}
where the prime denotes the derivative with respect to $\eta$ and $\cH=a'/a$
is the conformal Hubble rate. 
Note that the energy transfer by Thomson scattering may be ignored
\footnote{More precisely, this approximation is correct up to the first order of 
cosmological perturbation and we will use only the energy conservation law 
at the first order to derive the evolution equation for magnetic fields.}~\cite{Maartens:2007, Kobayashi:2007wd}.
Thus, the equations of motion governing the present 3-fluid system are given by 
\begin{eqnarray}
\nabla_{\nu}T_{(\gamma )i}{}^{\nu} &=& \kappa^{\gamma p}_i + \kappa^{\gamma e}_i , \label{ph}\\
\nabla_{\nu}T_{(p)i}{}^{\nu} &=& en_pE_i + \kappa^{pe}_i + \kappa^{p\gamma }_i , \label{pr}\\
\nabla_{\nu}T_{(e)i}{}^{\nu} &=& -en_eE_i  + \kappa^{ep}_i + \kappa^{e\gamma }_i , \label{el}
\end{eqnarray}
where we neglected the Lorentz force from the magnetic field because it will give rise to
higher order contributions.


To proceed, it is convenient to define
the center of mass velocity and the relative velocity as
\begin{eqnarray}
v_{(b)i} &:=& \frac{v_{(p)i}+\beta v_{(e)i}}{1+\beta},\\
\delta v_{(IJ)i} &:=& v_{(I)i} - v_{(J)i},
\end{eqnarray}
where $\beta:= m_e/m_p$. Hereafter we will assume charge neutrality:
$n_p=n_e (=:n)$ in the rest frame of charged particles. This will be 
justified because the velocity difference between protons and electrons, 
$\delta v_{(pe)i}$, is ignored compared to $\delta v_{(\gamma b)i}$ in the situation which the Coulomb interaction 
is very strong. 

Now the equation of motion for the photon fluid reads
\begin{widetext}
\begin{eqnarray}
&
\frac{4}{3}\Biggl\{
\rho_\gamma'(v_{(\gamma)i}+\chi_i) +\rho_\gamma(v_{(\gamma)i}+\chi_i)' 
   +4\cH\rho_\gamma(v_{(\gamma)i}+\chi_i) +\frac{1}{4}\dd_i\rho_\gamma
 - \left(\phi+2\cR\right)\left[\rho_\gamma v_{(\gamma)i}'
          + \rho_\gamma' v_{(\gamma)i}+4\cH\rho_\gamma v_{(\gamma)i}\right]
 \nonumber\\
&\qquad\qquad
 - 5\rho_\gamma\cR'v_{(\gamma)i}
 + \rho_\gamma\dd_i\left(\phi-\phi^2\right) 
 + \rho_\gamma\dd_j\left(v_{(\gamma)i}v_{(\gamma)}^j\right) \Biggr\}
=-\frac{4}{3}a\sigma_T n\rho_\gamma(1-2\cR)
  \Biggl[\left(1+\beta^2\right)\delta v_{(\gamma b)i} + \frac{1-\beta^3}{1+\beta}\delta v_{(pe)i}\Biggr] .
\label{gamma}
\end{eqnarray}
The equations of motion for the charged particles~(\ref{pr}) and~(\ref{el}) are combined to give
\begin{eqnarray}
&m_p(1+\beta)\Biggl\{
n'(v_{(b)i}+\chi_i) +n(v_{(b)i}+\chi_i)' +4\cH n(v_{(b)i}+\chi_i)
 -\left(\phi+2\cR\right)\left[n v_{(b)i}'+n' v_{(b)i}+4\cH n v_{(b)i}\right] -5n\cR'v_{(b)i}
\nonumber \\ &\qquad
 +n\dd_i\left(\phi-\phi^2\right) 
 +\frac{n}{1+\beta}\dd_j\left(v_{(p)i}v_{(p)}^j+\beta v_{(e)i}v_{(e)}^j\right)\Biggr\}
=\frac{4}{3}a\sigma_T n\rho_\gamma(1-2\cR)
  \Biggl[\left(1+\beta^2\right)\delta v_{(\gamma b)i} + \frac{1-\beta^3}{1+\beta}\delta v_{(pe)i}\Biggr] .
\label{b}
\end{eqnarray}

The equations of motion for the charged particles can be combined in a different way.
The manipulation $m_en_e\times\,$(\ref{pr}) $-m_pn_p\times\,$(\ref{el}) leads
to another independent equation: 
\begin{eqnarray}
&
m_em_pn \Bigg\{  
       \left(n\delta v_{(pe)i}\right)' + 4\cH n\delta v_{(pe)i}
      -\left(\phi+2\cR\right)
          \left[\left(n\delta v_{(pe)i}\right)' + 4\cH n\delta v_{(pe)i}\right]  
 -5\cR'n\delta v_{(pe)i}+n\dd_j\left(v_{(p)i}v_{(p)}^j-v_{(e)i}v_{(e)}^j\right)\Bigg\}
\nonumber\\&
= en^2m_p\left(1+\beta\right)E_i
     -\frac{4}{3}a\sigma_T n^2m_p\rho_\gamma(1-2\cR) 
  \left[\left(1-\beta^3\right)\delta v_{(\gamma b)i} + \frac{1+\beta^4}{1+\beta}\delta v_{(pe)i}\right] 
-e^2m_p n^3(1+\beta)\eta_C (1-2\cR)\delta v_{(pe)i}.
\label{p-e}
\end{eqnarray}
\end{widetext}
Since the Coulomb interaction is very strong in the regime we consider, 
we may take
the limit $\delta v_{(\gamma b)i}\gg \delta v_{(pe)i}\to0$. 
Then Eq.~(\ref{p-e}) simply reduces to
\begin{equation}
E_i = \frac{1-\beta^3}{1+\beta}\frac{4\sigma _T}{3e}a\rho_\gamma
       (1-2\cR)\delta v_{(\gamma b)i} .\label{ohmslaw}
\end{equation}
This may be regarded as ``Ohm's law'' in some sense. 
In the standard Ohm's law, the electric field is proportional 
to the electric current density $\sim e \delta v_{(pe)i}$.  
It is reminded that the contribution from the 
electric current gives us the diffusion term in the evolution equation for the 
magnetic field, and then the source for the magnetic field cannot be 
induced. 
However, the electric field is proportional to $\delta v_{(\gamma b)i}$ 
in the above formula.  The current in our ``Ohm's law" is originated from the 
velocity difference between protons and electrons through 
the interaction with photons. Indeed, if one takes the same mass limit 
of $\beta \to 1$ (though it is not realized in the nature), the electric 
field cannot be generated.

\subsection{Magnetic fields}

Let us turn to the evolution of the electromagnetic fields.
The Bianchi identities, $\nabla_{[\lambda}F_{\mu\nu]}= 0 $,
yield the induction equation
\begin{eqnarray}
\left(a^3B^i\right)' 
=-\epsilon^{ijk}\dd_j\left[a\left(1+\phi\right)E_k\right]
      -\epsilon^{ijk}\left(av_jE_k\right)'.
\end{eqnarray}
Using ``Ohm's law"~(\ref{ohmslaw}), this can be written as
\begin{widetext}
\begin{eqnarray}
\left(a^3B^i\right)' 
&=& -\frac{1-\beta^3}{1+\beta}\frac{4\sigma _T}{3e}
    \times \epsilon^{ijk}a^2\left[
    \dd_j\left(\rho_\gamma\delta v_{(\gamma b)k}\right) 
     + \rho_\gamma\dd_j(\phi-2\cR)\delta v_{(\gamma b)k}
      +\frac{1}{a^2}\left(\rho_\gamma v_ja^2\delta v_{(\gamma b)k}\right)'
\right].\label{Mag1}
\end{eqnarray}
\end{widetext}
It is now clear from Eq.~(\ref{Mag1}) that we need
the velocity difference between the photons and the charged particles, $\delta v_{(\gamma b)i}$,
to evaluate the generated magnetic field. 
The velocity difference can be computed as follows.
The manipulation
$(4\rho_\gamma/3)^{-1}\times (\ref{gamma}) - [m_p\left(1+\beta\right)n]^{-1}\times (\ref{b})$
leads to
\begin{widetext}
\begin{eqnarray}
& & 
\frac{\rho_\gamma'}{\rho_\gamma}\left(v_{(\gamma)i}+\chi_i\right) 
     -\frac{n'}{n}\left(v_{(b)i}+\chi_i\right) 
 +\left(\delta v_{(\gamma b)i}\right)' +4\cH\delta v_{(\gamma b)i} 
-\left(\phi+2\cR\right)\left[\frac{\rho_\gamma'}{\rho_\gamma}v_{(\gamma)i}
-\frac{n'}{n}v_{(b)i}
  +\left(\delta v_{(\gamma b)i}\right)'+4\cH\delta v_{(\gamma b)i}\right]
\nonumber\\&&\qquad\qquad\qquad
- 5\cR'\delta v_{(\gamma b)i}
+ \frac{1}{4}\frac{\dd_i\rho_\gamma}{\rho_\gamma}
+\dd_j\left(v_{(\gamma)i}v_{(\gamma)}^j\right)
 -\frac{1}{1+\beta}\dd_j\left(v_{(p)i}v_{(p)}^j+\beta v_{(e)i}v_{(e)}^j\right) 
=-\alpha (1-2\cR)\delta v_{(\gamma b)i}
\label{evo-v},
\end{eqnarray}
\end{widetext}
where $\alpha$ is defined as 
\begin{equation}
\alpha := \frac{1+\beta^2}{1+\beta}(1+R)\frac{4a\sigma_T \rho_\gamma }{3m_p} 
\left(
=\frac{4\beta(1+\beta^2)}{3(1+\beta)}(1+R)\frac{1}{\tau_T}
\right)
\end{equation}
with
\begin{equation}
R:=\frac{3m_p\left(1+\beta\right)n}{4\rho_\gamma }
\end{equation}
and
\begin{equation}
\tau_T:=\frac{m_e}{a \sigma_T \rho_\gamma}
\end{equation}
In deriving the above equation we again used $\delta v_{(pe)i}\ll \delta v_{(\gamma b)i}$.
Thus, our remaining task is to solve Eq.~(\ref{evo-v}) for $\delta v_{(\gamma b)i}$.

%
%
\section{Tight coupling approximation}

We are to solve Eq.~(\ref{evo-v}) using the tight coupling
approximation (TCA)~\cite{Peebles:1970ag}.
In this approximation
the time scale of Thomson scattering ($\tau_T = m_e/a\sigma_T\rho_\gamma$) 
is assumed to be much 
smaller than the wavelengths of the perturbations ($k^{-1}$). Thus, the small 
expansion parameter of the TCA is $k \tau_T$, which is dimensionless.
During the pre-recombination era, photons, protons, and electrons are strongly coupled 
via Thomson scattering, and hence the TCA will be a good approximation.

At zeroth order in the TCA, all fluid components have the same velocity $v_i$ and
the density 
fluctuations are adiabatic.
Following Ref.~\cite{Takahashi:2007ds}, we define the deviation from 
the adiabatic distribution for baryons by
\begin{eqnarray}
n_b &= \bar{n}_b\left(1+\Delta_b\right).
\end{eqnarray}
Then, we expand various quantities such as
$\Delta_b$ and $v_{(I)i}$ in terms of the tight coupling parameter $k \tau_T $:
\begin{eqnarray}
& & \Delta_b = \Delta_b^{(I)} + \Delta_b^{(II)} +\cdots, \\
& & v_{(\gamma)i} = v_i, \\
& & v_{(b)i} = v_i+v_{(b)i}^{(I)}+v_{(b)i}^{(II)}+\cdots , \\
& & \delta v_{(\gamma b)i} = \delta v_{(\gamma b)i}^{(I)} 
 + \delta v_{(\gamma b)i}^{(II)}+\cdots,
\end{eqnarray}
where $v_i$ is the common velocity of photons and baryons 
in the tight coupling limit.
Our notation is that Roman and Arabic numerals stand for the order of TCA and 
that in cosmological perturbation theory, respectively.
Here, we adopt the photon frame, so that 
$\Delta_\gamma^{(I)}=\Delta_\gamma^{(II)}=0$. 
In other words, the quantities associated with photons give the ``background''
in the TCA.
Note that we consider cosmological perturbation theory and the TCA simultaneously. 
The following analysis includes cosmological perturbations up to second-order
and the tight coupling expansion
up to TCA(II), where TCA($n$) denotes the tight coupling approximation at $n$-th order.
The notation $X^{(n, m)}$ indicates that $X$ is
a $m$-th order quantity in cosmological perturbation theory
at TCA($n$).

At each order in the TCA $(n=I, II)$, Eq.~(\ref{Mag1}) reduces to
\begin{eqnarray}
\left[\left(a^3 B^i\right)'\right]^{(n)} = -\frac{1-\beta^3}{1+\beta}\frac{4\sigma_T}{3e}a^2\bar{\rho}_{\gamma}^{(0)}
\left(\tilde S^i_{\;B}\right)^{(n)},
\end{eqnarray}
where
\begin{eqnarray}
\left(\tilde S^i_{\;B}\right)^{(n)}&=&
\epsilon^{ijk}\Biggl[\partial_j \delta v_{(\gamma b)k}^{(n,2)}+ \frac{1}{\bar \rho_\gamma^{(0)}} \partial_j 
\left(\delta \bar \rho_\gamma^{(1)} \delta v_{(\gamma b)k}^{(n,1)}\right)
\nonumber\\&&\quad
+ \partial_j (\phi^{(1)} - 2\cR^{(1)})\delta v_{(\gamma b)k}^{(n,1)} 
\nonumber\\&&\quad
+ \frac{1}{a^2 \bar \rho_\gamma^{(0)}}
   \left(a^2 \bar{\rho}_\gamma^{(0)}v_j^{(1)} 
   \delta v_{(\gamma b)k}^{(n,1)}\right)' \Biggr]\label{MagI+II}
\end{eqnarray}
We shall now evaluate the source term $\left(\tilde S^i_{\;B}\right)^{(n)}$.



\subsection{TCA(0)}

\subsubsection{Energy conservation}

At $(0,0)$ order, Eq.(\ref{En1}) reads 
\begin{eqnarray}
\left(\bar{\rho}_{\gamma}^{(0)}\right)' + 4\cH\bar{\rho}_\gamma^{(0)} =0,
\quad
\left(\bar{n}^{(0)}\right)' + 3\cH\bar{n}^{(0)} =0. 
\label{ene0}
\end{eqnarray}
At next order in cosmological perturbation theory, i.e., at $(0, 1)$ order, we similarly have
\begin{eqnarray}
-\left(\delta\bar{\rho}_\gamma^{(1)}\right)'-4\cH\delta\bar{\rho}^{(1)}_\gamma
 &+&4\bar{\rho}_\gamma^{(0)}\left(\cR^{(1)}\right)'
\nonumber\\&&
   -\frac{4}{3}\bar{\rho}_\gamma^{(0)}\partial_iv^{i(1)} =0 ,\\
-\left(\delta \bar{n}^{(1)}\right)'-3\cH\delta \bar{n}^{(1)}
 &+&3\bar{n}^{(0)}\left(\cR^{(1)}\right)'
\nonumber\\&&
 -\bar{n}^{(0)}\partial_iv^{i(1)} =0.
\end{eqnarray}
These equations can be put into
\begin{eqnarray}
 \left(\frac{\delta\bar{\rho}_\gamma^{(1)}}{\bar{\rho}^{(0)}_\gamma}\right)'
&=& 4\left(\cR^{(1)}\right)' -\frac{4}{3}\dd_i v^{i(1)} ,\quad
\nonumber\\
 \left(\frac{\delta \bar{n}^{(1)}}{\bar{n}^{(0)}}\right)'
&=& 3\left(\cR^{(1)}\right)'-\dd_i v^{i(1)} 
\label{ene1},
\end{eqnarray}
implying that
\begin{equation}
\left(\frac{\delta\bar{\rho}_\gamma^{(1)}}{\bar{\rho}^{(0)}_\gamma}\right)'
=\frac{4}{3}\left(\frac{\delta \bar{n}^{(1)}}{\bar{n}^{(0)}}\right)'. 
\end{equation}

\subsubsection{Vorticity}
We show that the vorticity vanishes at TCA(0). Using Eq. (\ref{ene0}),
the momentum conservations for photons and baryons at (0,1) order are
\begin{eqnarray}
  (v_i^{(1)})'
+ \frac{1}{4}\frac{\dd_i\delta\bar{\rho}_\gamma^{(1)}}{\bar{\rho}_\gamma^{(0)}} 
+ \dd_i\phi^{(1)}
&=&0 \label{gam1},
\nonumber\\
  (v_i^{(1)})'
+ \cH v_i^{(1)}
+ \dd_i\phi^{(1)}
&=&0 .\label{bar1}
\end{eqnarray}
We eliminate $(v_i^{(1)})'$ from the above equations and obtain the relation
\begin{equation}
\cH v_i^{(1)}=\frac{1}{4}\frac{\dd_i\delta\bar{\rho}_\gamma^{(1)}}
 {\bar{\rho}_\gamma^{(0)}}
\label{veq1}
.
\end{equation}
At (0,2) order, we eliminate similarly $(v_i^{(2)}+\chi_i^{(2)})'$. 
The momentum conservation gives 
\begin{eqnarray}
&&   (v_i^{(2)}+\chi_i^{(2)})
- \left(\phi^{(1)}+2\cR^{(1)}\right)v_i^{(1)}
\nonumber\\& &
- \frac{1}{4\cH}\left(
   -\frac{\delta\bar{\rho}_\gamma^{(1)}}{\bar{\rho}_\gamma^{(0)2}}
   \delta\bar{\rho}_\gamma^{(1)}+\frac{\dd_i\delta\bar{\rho}_\gamma^{(2)}}
   {\bar{\rho}_\gamma^{(0)}}\right)
\nonumber\\& &
+ \frac{1}{4\cH^2}\left(-(\cR^{(1)})' + \frac{1}{3}\dd_\ell v^{(1)\ell}\right)
   \frac{\dd_i\delta\bar{\rho}_\gamma^{(1)}}{\bar{\rho}_\gamma^{(0)}}
=0.
\end{eqnarray}
Taking the curl of the above equation and using Eq.~(\ref{veq1}), we can
show that $\omega^{(2)i}=0$.
It indicates that the vorticity is always zero at TCA(0).


\subsection{TCA(I)}

\subsubsection{Energy conservation}

To compute $\delta v_{(\gamma b)i}^{(II,2)}$,
we evaluate $\Delta_b^{(I,1)}$ using the energy conservation.
At $(I, 1)$ order, Eq.~(\ref{En1}) reduces to 
\begin{eqnarray}
-\left(\bar{n}^{(0)}\Delta_b^{(I,1)}\right)' -3\cH\bar{n}^{(0)}\Delta_b^{(I,1)}
 -\bar{n}^{(0)}\dd_iv_{(b)}^{(I,1)i}=0,
\end{eqnarray}
and hence
\begin{eqnarray}
\left(\Delta_b^{(I,1)}\right)' +\dd_iv_{(b)}^{(I,1)i}=0 ,
\end{eqnarray}
which is integrated to give 
\begin{eqnarray}
\Delta_b^{(I,1)} =\int^{\eta} d\eta\,\dd_i\delta v_{(\gamma b)}^{(I,1)i}.
\end{eqnarray}

\subsubsection{Vorticity evolution equation}
To derive the vorticity evolution equation, we take the curl of
the total momentum conservation. The total momentum conservation is
\begin{eqnarray}
& &
  \left(\frac{4}{3}\rho_\gamma v_{(\gamma)i}+\rho_b v_{(b)i} 
+ \rho_T\chi_i\right)' 
+ 4\cH\left(\frac{4}{3}\rho_\gamma v_{(\gamma)i} 
+ \rho_b v_{(b)i}\right)
\nonumber\\&&~~
+4\cH\rho_T\chi_i
- 5\cR'\left(\frac{4}{3}\rho_\gamma v_{(\gamma)i}+\rho_b v_{(b)i}\right) 
+ \frac{1}{3}\dd_i \rho_\gamma 
\nonumber \\& &~
-\left(\phi+2\cR\right)
\left[
\left(\frac{4}{3}\rho_\gamma v_{(\gamma)i}+\rho_b v_{(b)i}\right)' 
\right.\nonumber\\ &&\qquad\qquad\qquad\qquad\quad
\left. 
+4\cH\left(\frac{4}{3}\rho_\gamma v_{(\gamma)i}+\rho_b v_{(b)i}\right)
\right]
\nonumber\\&& 
+ \rho_T\dd_i(\phi-\phi^2) 
+\sum_I\rho_I(1+w_I)\dd_j\left(v_{(I)i} v_{(I)}^j\right) = 0 ,
\label{Total1.eq}
\end{eqnarray}
where $\rho_b =m_p(1+\beta)n$. Note that the scattering terms are
cancelled out. 
At TCA(I), this equation is simplified to 
\begin{eqnarray}
  \left[\rho_T(v_i+\chi_i)\right]' &+& 4\cH\rho_T(v_i+\chi_i)
+ \rho_T\dd_i(\phi-\phi^2) 
\nonumber\\
&+& \frac{1}{3}\dd_i \rho_\gamma  
 -\left(\phi+2\cR\right)\left[\left(\rho_T v_i\right)'+4\cH\rho_T v_i\right]
\nonumber\\
&-& 5\rho_T\cR'v_i +\rho_T\dd_j\left(v_i v^j\right) = 0 ,\label{bgv}
\end{eqnarray}
where $\rho_T$ is defined by 
\begin{eqnarray}
\rho_T := \frac{4}{3}\rho_\gamma + \rho_b =\frac{4}{3}\rho_\gamma (1+R). 
\label{rhot}
\end{eqnarray}

At $(0, 1)$ order, Eq.~(\ref{bgv}) reads 
\begin{eqnarray}
\left(v_i^{(1)}\right)' +\frac{\left(\bar{\rho}_T^{(0)}\right)'}{\bar{\rho}_T^{(0)}}v_i^{(1)}
+4\cH
v_i^{(1)}+\frac{1}{3}\frac{\dd_i\delta\bar{\rho}^{(1)}_\gamma}{\bar{\rho}^{(0)}_T} 
\nonumber\\
+\dd_i\phi^{(1)}= 0,
\end{eqnarray}
or, equivalently,
\begin{equation}
\left(v_i^{(1)}\right)' +\frac{\cH \bar{R}^{(0)}}{1+\bar{R}^{(0)}}v_i^{(1)}
=-\frac{1}{4\left(1+\bar{R}^{(0)}\right)}\frac{\dd_i\delta\bar{\rho}^{(1)}_\gamma}{\bar{\rho}^{(0)}_\gamma}-\dd_i\phi^{(1)}.\label{bgv1} 
\end{equation}
Taking the curl of Eq.~(\ref{bgv}), we arrive at the vorticity
conservation equation (for the photon fluid):
\begin{eqnarray}
\left(2a^2\bar{\rho}_T^{(0)}\omega^{i(2)}\right)'+8a^2\cH\bar{\rho}_T^{(0)}\omega^{i(2)} = 0, \label{vol0}
\end{eqnarray}
where we used Eq. (\ref{bgv1}). 
It is important to note here that
there is no source for the vorticity at TCA(I).
Therefore, we may assume that $\omega^{(2)}_i=0$ at TCA(I).


\subsubsection{Derivation of $\delta v_{(\gamma b)i}^{(I,1)}$ and $\delta v_{(\gamma b)i}^{(I, 2)}$}

At $(I, 1)$ order,  Eq.~(\ref{evo-v}) reads
\begin{eqnarray}
\left[\frac{\left(\bar{\rho}_\gamma^{(0)}\right)'}{\bar{\rho}_\gamma^{(0)}}
        -\frac{\left(\bar{n}^{(0)}\right)'}{\bar{n}^{(0)}}\right]v_i^{(1)}
+\frac{1}{4}\frac{\dd_i \delta\bar{\rho}_\gamma^{(1)}}{\bar{\rho}_\gamma^{(0)}}
= -\bar{\alpha}^{(0)}\delta v_{(\gamma b)i}^{(I,1)},
\end{eqnarray}
and so we have
\begin{equation}
\delta v_{(\gamma b)i}^{(I,1)} 
= \frac{1}{\bar{\alpha}^{(0)}}\left[\cH v_i^{(1)}
    -\frac{1}{4}\frac{\dd_i
    \delta\bar{\rho}_\gamma^{(1)}}{\bar{\rho}_\gamma^{(0)}}\right], 
\label{dv11}
\end{equation}
where 
\begin{equation}
\bar \alpha^{(0)}=\frac{(1+\beta^2)(1+\bar R^{(0)})}{1+\beta}
  \frac{4a\sigma_T\bar{\rho}_\gamma^{(0)}}{3m_p} .\label{baralpha0} 
\end{equation}

At next order in cosmological perturbation theory, i.e., at $(I, 2)$
order, we have 
\begin{widetext}
\begin{eqnarray}
&&\Biggl[\frac{\left(\delta\bar{\rho}_\gamma^{(1)}\right)'}{\bar{\rho}_\gamma^{(0)}}
 -\frac{\delta\bar{\rho}_\gamma^{(1)}}{\bar{\rho}_\gamma^{(0)}}
  \frac{\left(\bar{\rho}_\gamma^{(0)}\right)'}{\bar{\rho}_\gamma^{(0)}}
 -\left\{\frac{\left(\delta\bar{n}^{(1)}\right)'}{\bar{n}^{(0)}}
          -\frac{\delta\bar{n}^{(1)}}{\bar{n}^{(0)}}
  \frac{\left(\bar{n}^{(0)}\right)'}{\bar{n}^{(0)}}\right\}
 -\left(\phi^{(1)}+2\cR^{(1)}\right)
   \left\{\frac{\left(\bar{\rho}_\gamma^{(0)}\right)'}{\bar{\rho}_\gamma^{(0)}}
        -\frac{\left(\bar{n}^{(0)}\right)'}{\bar{n}^{(0)}}\right\}
\Biggr]v_i^{(1)}
\nonumber\\&&\;
+\left[\frac{\left(\bar{\rho}_\gamma^{(0)}\right)'}{\bar{\rho}_\gamma^{(0)}}
        -\frac{\left(\bar{n}^{(0)}\right)'}{\bar{n}^{(0)}}\right]
\left(v_i^{(2)}+\chi^{(2)}_i\right)
-\frac{1}{4}\left[
\frac{\delta\bar{\rho}_\gamma^{(1)}}{\bar{\rho}_\gamma^{(0)}}
   \frac{\dd_i \delta\bar{\rho}_\gamma^{(1)}}{\bar{\rho}_\gamma^{(0)}}
-\frac{\dd_i \delta\bar{\rho}_\gamma^{(2)}}{\bar{\rho}_\gamma^{(0)}}
\right]
=- \bar{\alpha}^{(0)}\delta v_{(\gamma b)i}^{(I,2)}
 + 2\cR^{(1)}\bar{\alpha}^{(0)}\delta v_{(\gamma b)i}^{(I,1)}
 - \bar{\alpha}^{(1)}\delta v_{(\gamma b)i}^{(I,1)},
\nonumber\\
\end{eqnarray}
where 
\begin{eqnarray}
\bar{\alpha}^{(1)} 
=\frac{1}{\Bigl(1+\bar{R}^{(0)}\Bigr)\bar{\rho}_\gamma^{(0)}}
 \bar{\alpha}^{(0)}  \Biggl[\delta\bar{\rho}_\gamma^{(1)}+\frac{3}{4}(m_p+m_e)\delta \bar{n}^{(1)}\Biggr]
=\frac{4+3\bar{R}^{(0)}}{4\Bigl(1+\bar{R}^{(0)}\Bigr)}
 \bar{\alpha}^{(0)}   \frac{\delta\bar{\rho}_\gamma^{(1)}}{\bar{\rho}_\gamma^{(0)}}. 
\end{eqnarray}
This can be solved for $\delta v_{(\gamma b)i}^{(I,2)}$ to give
\begin{eqnarray}
\delta v_{(\gamma b)i}^{(I,2)} 
&=& \frac{1}{\bar{\alpha}^{(0)}}\Bigg[
    \cH\left(v_i^{(2)}+\chi_i^{(2)}-\frac{\bar{\alpha}^{(1)}}{\bar{\alpha}^{(0)}}v_i^{(1)}\right) 
  -\left\{\left(\cR^{(1)}\right)' -\frac{1}{3}\dd_\ell v^{\ell(1)}\right\}v_i^{(1)}
\nonumber \\ &&\qquad\quad
 -\cH\phi^{(1)}v_i^{(1)}
   -\frac{1}{4}\left\{
    \frac{\dd_i \delta\bar{\rho}_\gamma^{(2)}}{\bar{\rho}_\gamma^{(0)}}  
   -\left(\frac{\delta\bar{\rho}_\gamma^{(1)}}{\bar{\rho}_\gamma^{(0)}}
     +\frac{\bar{\alpha}^{(1)}}{\bar{\alpha}^{(0)}}\right)
       \frac{\dd_i \delta\bar{\rho}_\gamma^{(1)}}{\bar{\rho}_\gamma^{(0)}}
\right\}
  - \frac{\cR^{(1)}}{2}\frac{\dd_i
    \delta\bar{\rho}_\gamma^{(1)}}{\bar{\rho}_\gamma^{(0)}}
\Bigg] .
\label{dv12}
\end{eqnarray}
\end{widetext}

\subsection{TCA(II)}

\subsubsection{Derivation of $\delta v_{(\gamma b)i}^{(II,1)}$ and $\delta v_{(\gamma b)i}^{(II, 2)}$}

Eq.~(\ref{evo-v}) at order $(I, 1)$ and $(I, 2)$ allows us to obtain the 
expression for $\delta v_{(\gamma b)i}^{(II, 1)}$ and $\delta v_{(\gamma b)i}^{(II, 2)}$, respectively.
At first 
order in cosmological perturbation theory, we have 
\begin{equation}
-\frac{\left(\bar{n}^{(0)}\right)'}{\bar{n}^{(0)}}v_{(b)i}^{(I,1)}
+\left(\delta v_{(\gamma b)i}^{(I,1)}\right)' +4\cH\delta v_{(\gamma b)i}^{(I,1)}
=-\bar{\alpha}^{(0)}\delta v_{(\gamma b)i}^{(II,1)},
\end{equation}
yielding
\begin{eqnarray}
\delta v_{(\gamma b)i}^{(II,1)}
= -\frac{1}{\bar{\alpha}^{(0)}}
\left[
  \left(\delta v_{(\gamma b)i}^{(I,1)}\right)' +\cH \delta v_{(\gamma b)i}^{(I,1)}
\right]\label{dv21}.
\end{eqnarray}
At second order in cosmological perturbation theory, we get
\begin{widetext}
\begin{eqnarray}
&&  \left(\delta v_{(\gamma b)i}^{(I,2)}\right)'
+ \cH\delta v_{(\gamma b)i}^{(I,2)}
- \left(\phi^{(1)}+2\cR^{(1)}\right)\Biggl[\left(\delta v_{(\gamma b)i}^{(I,1)}
  \right)' + \cH\delta v_{(\gamma b)i}^{(I,1)}\Biggr]
- 2\left(\cR^{(1)}\right)' \delta v_{(\gamma b)i}^{(I,1)}
\nonumber\\&&
+ \left(\dd_\ell v^{(1)}_i\right)\delta v_{(\gamma b)}^{(I,1)\ell}
+ \left(\dd_\ell\delta v_{(\gamma b)i}^{(I,1)}\right)v^{(1)\ell}
=
- \bar{\alpha}^{(0)}\delta v_{(\gamma b)i}^{(II,2)}
+ 2\cR^{(1)}\bar{\alpha}^{(0)}\delta v_{(\gamma b)i}^{(II,1)}
- \bar{\alpha}^{(1)}\delta v_{(\gamma b)i}^{(II,1)}
- \alpha^{(I,1)}\delta v_{(\gamma b)i}^{(I,1)}.
\end{eqnarray}
From this we obtain
\begin{eqnarray}
&&\delta v_{(\gamma b)i}^{(II,2)}
=  -\frac{1}{\bar \alpha^{(0)}}\Bigl[ (\delta v_{(\gamma b)i}^{(I,2)})'+\cH 
\delta v_{(\gamma b)i}^{(I,2)}  \Bigr]
+   \frac{\bar \alpha^{(1)}}{(\bar \alpha^{(0)})^2}
     \Bigl[ \left(\delta v_{(\gamma b)i}^{(I,1)}\right)'
     + \cH \delta v_{(\gamma b)i}^{(I,1)}\Bigr]
-\frac{\alpha^{(I,1)}}{\bar \alpha^{(0)}}\delta v_{(\gamma b)i}^{(I,1)}
\nonumber\\&&\qquad
+\frac{1}{\bar \alpha^{(0)}}
\Biggl[ 
 \phi^{(1)}\Bigl\{ (\delta v_{(\gamma b)i}^{(I,1)})'+\cH 
 \delta v_{(\gamma b)i}^{(I,1)}  \Bigr\}
 +2(\cR^{(1)})' \delta v_{(\gamma b)i}^{(I,1)}
-\dd_\ell v_i^{(1)} \delta v_{(\gamma b)}^{(I,1)\ell}
-\dd_\ell \delta v_{(\gamma b)i}^{(I,1)}v^{(1)\ell}
\Biggr],\label{dv22}
\end{eqnarray}
\end{widetext}
where we used Eq.~(\ref{dv21}) and note that
\begin{eqnarray}
\frac{\alpha^{(I,1)}}{\bar{\alpha}^{(0)}}
=\frac{\bar{R}^{(0)}}{1+\bar{R}^{(0)}}\Delta_b^{(I,1)}
=\frac{\bar{R}^{(0)}}{1+\bar{R}^{(0)}}\int d\eta\,\dd_i\delta v_{(\gamma b)}^{(I,1)i} .\label{alphaI1}
\end{eqnarray}

\subsubsection{Vorticity evolution equation}
In order to derive the vorticity evolution equation at TCA(II), 
we need to consider Eq.(\ref{Total1.eq}) at TCA(I). At 
$(I, 1)$ order, 
Eq.~(\ref{Total1.eq}) reads 
\begin{eqnarray}
&&(v_i^{(1)})' = 
- \frac{\cH\bar{R}^{(0)}}{1+\bar{R}^{(0)}}v_i^{(1)} 
- \frac{1}{4(1+\bar{R}^{(0)})}\frac{\dd_i\delta\bar{\rho}_\gamma^{(1)}}
    {\bar{\rho}_\gamma^{(0)}}
\nonumber\\&&\qquad\quad
- \dd_i\phi^{(1)}
+ \frac{\cH\bar{R}^{(0)}}{1+\bar{R}^{(0)}}\left[
   (\delta v_{(\gamma b)i}^{(I,1)})'+\cH\delta v_{(\gamma b)i}^{(I,1)}
\right],
\label{vpi}
\end{eqnarray}
where we used Eqs.~(\ref{ene0}) and~(\ref{rhot}).
We take the curl of Eq.~(\ref{Total1.eq}) at $(I, 2)$ order,
so that we obtain 
\begin{eqnarray}
(a^2 \omega^{(2)i})'&+&\frac{\cH\bar R ^{(0)}}{1+\bar R ^{(0)}}a^2
 \omega^{(2)i}
=
\nonumber\\ &&
\frac{1}{2(1+\bar R^{(0)})} \epsilon^{ijk}
 \dd_j\alpha^{(I,1)}\delta v^{(I,1)}_{(\gamma b)k}
\label{vol2}
\end{eqnarray}
where we used Eq.~(\ref{vpi}) and
\begin{eqnarray}
&&\epsilon^{ijk}\dd_j\delta v_{(\gamma b)k}^{(I,2)}
=-\frac{2a^2\cH}{\bar \alpha^{(0)}}\omega^{(2)i}
+ 2\epsilon^{ijk}\dd_j\cR^{(1)}\delta v_{(\gamma b)k}^{(I,1)}
\nonumber\\&&\qquad\qquad
- \epsilon^{ijk}v_j^{(1)}\left[(\delta v_{(\gamma b)k}^{(I,1)})'
   +\cH\delta v_{(\gamma b)k}^{(I,1)}\right]
\label{eq71}
.
\end{eqnarray}
Since the right hand side in Eq. (\ref{vol2}) contain the source term for 
the vorticity at TCA(II), the vorticity can be generated at this order.

\subsection{Generation of magnetic fields}

To evaluate the magnetic fields we substitute 
$\delta v_{(\gamma b)i}^{(n, m)}$
[i.e., Eqs.~(\ref{dv11}), (\ref{dv12}), (\ref{dv21}), and (\ref{dv22})]
into the source term for magnetic fields [i.e., Eq.~(\ref{MagI+II})].
It is easy to see that 
the source term at TCA(I) is given by  
\begin{eqnarray}
(\tilde S_B^i)^{(I)} &=& -\frac{2a^2\cH}{\bar \alpha^{(0)}}\omega^{(2)i} 
\end{eqnarray}
where we used Eqs.(\ref{bgv1}), (\ref{eq71}).
This turns out to be proportional to the vorticity of the photon fluid. Indeed, we find 
\begin{eqnarray}
[(a^3B^i)']^{(I)}=2\frac{1-\beta^3}{1+\beta}\frac{4\sigma_T}{3e} 
a^4 \bar{\rho}_\gamma^{(0)}\frac{\cH}{\bar \alpha^{(0)}}
\omega^{(2)i}. 
\end{eqnarray}
If one assumes a vanishing initial vorticity [see the argument around Eq.~(\ref{vol0})], 
magnetic fields cannot be produced at TCA(I). 

The source term which we consider up to TCA(II) is expressed as 
\begin{eqnarray}
(\tilde S_B^i)&=&(\tilde S_B^i)^{(I)}+(\tilde S_B^i)^{(II)}
\nonumber\\ 
&=&
-\frac{2a^2\cH}{\bar \alpha^{(0)}}
\omega^{(2)i}
-\epsilon^{ijk}\frac{\dd_j \alpha^{(I,1)}}{\bar \alpha^{(0)}}
\delta v_{(\gamma b)k}^{(I,1)},
\label{s2}
\end{eqnarray}
where $\bar \alpha^{(0)}$, $\alpha^{(I,1)}$, and $\delta v_{(\gamma b)i}^{(I, 1)}$ are given by 
Eqs.~(\ref{baralpha0}), (\ref{alphaI1}), and (\ref{dv11}), respectively.
Note in passing that
$(\delta v_{(\gamma b)i}^{(I, 1)})'$ is computed as 
\begin{eqnarray}
(\delta v_{(\gamma b)i}^{(I, 1)})' &=&
\frac{\cH (3+2 \bar R^{(0)})}{1+\bar R^{(0)}}\delta v_{(\gamma b)i}^{(I,1)}
\nonumber\\&&
+\frac{1}{\bar \alpha^{(0)}}
\Biggl[ 
(\cH v_i^{(1)})'-\frac{1}{4}\partial_i \Bigl( \frac{\delta \bar \rho_\gamma^{(1)}}{\bar \rho_\gamma^{(0)}} 
\Bigr)' \Biggr]. 
\end{eqnarray}
Hence, we get 
\begin{eqnarray}
&&(a^3B^i)'=
\frac{1-\beta^3}{1+\beta}\frac{4\sigma_T}{3e} 
a^2 \bar{\rho}_\gamma^{(0)}
\times
\nonumber\\&&
\times
\left[
\frac{2a^2\cH}{\bar \alpha^{(0)}}
\omega^{(2)i}
+ \epsilon^{ijk}\frac{\bar{R}^{(0)}}{1+\bar{R}^{(0)}}
\dd_j\Delta_b^{(I,1)}\delta v_{(\gamma b)k}^{(I,1)}
\right].
\label{MagneII}
\end{eqnarray}
where we used Eq.(\ref{alphaI1}).
Intrinsically second-order perturbations appear only 
as $\omega^{(2)i}$ in the source term,
which is governed by Eq. (\ref{vol2}). 
We can check that Eq. (\ref{MagneII}) is the same as the result obtained 
in Ref. \cite{Kobayashi:2007wd}. Note that the vorticity is the general
relativistic effect because it vanishes in the limit of $\cH\to 0$. 
Taking the Newton limit ($\cH \to 0$), only the last term in the right-hand 
side of Eq. (\ref{MagneII}) remains. At first glance, 
this result is not the same as one obtained in Ref.~\cite{Takahashi:2007ds} 
where the tight coupling approximation was employed in Newtonian theory. 
This is not surprising because photon's rest frame is not 
used as the background of the tight coupling approximation in 
Ref.~\cite{Takahashi:2007ds}. 
Performing an appropriate coordinate transformation,
therefore, we can show that our result agrees 
exactly with \cite{Takahashi:2007ds}. 

Now we are ready to give an estimate of the magnetic field 
generated from second-order 
cosmological perturbations at the recombination epoch. 
For clarity and simplicity, we focus on the magnetic fields on the
horizon scales ($k \sim \cH $).
We find $\cR \sim \delta$ by using the first order Einstein equations
and $v \sim (\cH/k)\delta \sim \delta$ by using Eq.~(\ref{ene1}) on the horizon scales,
where $\delta \equiv \delta\bar{\rho}_\gamma^{(1)}/\bar{\rho}_\gamma^{(0)}$. 
Finally, the magnetic field 
on the horizon scales is estimated to be 
\begin{eqnarray}
B \sim \sqrt{B^iB_i} \sim
\frac{a^2m_p^2 k^2}{e\sigma_T (\bar{\rho}_{\gamma}^{(0)}a^4)}
\delta^2 \sim 10^{-29}\;{\rm Gauss},
\label{oe}
\end{eqnarray}
where we used $a \sim 10^{-3}$, 
$k \sim  (100\;{\rm Mpc})^{-1} \sim 0.64 \times10^{-40}\;{\rm GeV}$,
$e \sim 0.3$, $m_p \sim 0.94\;{\rm GeV}$, 
$\sigma_T \sim 1.7 \times 10^3\; ({\rm GeV})^{-2}$, and 
$\bar{\rho}_{\gamma}^{(0)}a^4 = \rho_{\gamma 0} \sim 2.1\times10^{-51} ({\rm GeV})^4$. This result is
consistent with the result in Refs.~\cite{Matarrese, Kobayashi:2007wd,
Ichiki2}.
According to \cite{Davis:1999bt}, this will be amplified enough to explain 
the present observed magnetic fields.

\section{Summary}

In this paper we have derived an analytic formula for the magnetic fields
generated from second-order cosmological perturbations
in the pre-recombination era.
Since first-order vector perturbations are not generated from inflation 
in the standard scenarios, we must go beyond the linear analysis
to generate magnetic fields.
Photons and charged particles are strongly coupled via Thomson scattering
within the cosmic plasma, and hence the system behaves almost as a single fluid.
In this single-fluid description magnetic fields are never generated,
and therefore
the tiny deviation from the single-fluid description is crucial here.
Using the tight coupling approximation (TCA)
to treat the small difference between photons and charged particles,
we have seen how magnetic fields are generated from cosmic inhomogeneities.
It was found that magnetic fields are not generated at 
first order in the TCA.
Therefore, we conclude that
magnetogenesis requires both
the second-order cosmological perturbations and the second-order TCA.  
The resultant magnetic 
fields are expressed in terms of the vorticity and the product of the first-order perturbations.
The latter can be computed by solving the linear Einstein equations,
while the former is governed by the vorticity evolution equation, with the source term
given by the product of the first order terms.
Finally, we gave an
order of magnitude estimate
for the magnetic fields generated on the horizon scales 
around the recombination epoch. The result was $B \sim 10^{-29}\;$Gauss.
This magnetic fields will be enhanced by the dynamo mechanism during and after galaxy formation 
to explain the present order of the magnitude of the observed magnetic fields 
\cite{Davis:1999bt}. Although our rough estimate suggests that the current 
generation mechanism can work even on smaller scales, say Mpc, more careful investigation is needed to 
make clear whether it indeed works in a wide range of spatial scales. 

This work is the general relativistic version of~\cite{Takahashi:2007ds},
including the effects of the cosmic expansion and metric perturbations.
The qualitative conclusion derived in this paper confirms the basic results obtained 
from the fully nonlinear, covariant analysis of Ref.~\cite{Kobayashi:2007wd}.
However, our formulation here is based on cosmological perturbation theory,
which will be more useful for actual calculations, e.g., of the spectrum of the magnetic fields.

We have not included the effect of anisotropic stresses of photon fluids
and the recombination process, though 
they will be equally important for magnetogenesis 
on small scales~\cite{Ichiki1, Ichiki2, Takahashi:2008gn}. 
Taking into account these effects and computing the power spectrum of the generated
magnetic fields require detailed numerical calculations, which are left for forthcoming publications.

\section*{Acknowledgements}

SK and TK thank Kei-ichi Maeda for useful comments.
TK is supported by the JSPS under Contract No.~19-4199. The work of 
TS was supported by Grant-in-Aid for Scientific Research from Ministry 
of Education, Science, Sports and Culture of Japan (Nos.~17740136,~17340075,~19GS0219, 
and 20540258). 

\appendix
\section{Consistency with Ref.\cite{Kobayashi:2007wd}}
We show that our result can be identified to that obtained 
by using the covariant approach \cite{Kobayashi:2007wd}.
The final result in Ref.~\cite{Kobayashi:2007wd} is
\begin{eqnarray}
\dot{B}^a&=&
- \frac{2}{3}\theta B^a+\sigma^a_{~b}B^b
\nonumber\\&&
- \left[\frac{m_p}{e}\frac{R}{4(1+R)^2}\frac{1-\beta^3}{1+\beta^2}\right]
   \varepsilon^{abc}D_b\Delta_{(1)}\frac{D_c\rho_\gamma}{\rho_\gamma}
\nonumber\\&&
- \left[\frac{m_p}{e}\frac{R}{2(1+R)}\frac{1-\beta^3}{1+\beta^2}
   \frac{\dot{\rho}_\gamma}{\rho_\gamma}\right]\omega^a.
\label{App1}
\end{eqnarray}
Here $\Delta_{(1)}$ is baryon's density fluctuation at TCA(I),
$\theta\equiv\nabla_a u^a$ is the volume expansion rate, 
$\sigma^a_{~b}\equiv
h^{~c}_ah^{~d}_b\nabla_{(c}u_{d)}-\frac{1}{3}\theta h_{ab}$ is the shear,
\begin{eqnarray}
\dot{A_a} := h_{~a}^{b}u^c\nabla_c A_b \label{App2}
\end{eqnarray}
for a tensor,
\begin{eqnarray}
\dot{f} := u^a\nabla_a f \label{App3}
\end{eqnarray}
for a scalar function and
\begin{eqnarray}
D_a := h_{~a}^{b}\nabla_b \label{App4}
\end{eqnarray}
where $u_a$ is photon's four velocity and $h_{ab}\equiv g_{ab}+u_au_b$ is the projection to
3-dimensional hypersurface normal to $u^a$.

Equation (\ref{App1}) is the nonlinear result.
For purpose of comparison we expand Eq.(\ref{App1})
in terms of cosmological perturbations up to second order.
We use the perturbed metric (\ref{metric}), and
photon's four velocity is defined 
by Eqs.(\ref{four1}) and (\ref{four2}).
Using our notation, 
the $i$ component of Eq.(\ref{App1}) is
\begin{eqnarray}
&&\dot{B}^{(2)i}
+ \frac{2}{3}\theta^{(0)} B^{(2)i}
=
\nonumber\\&&
- \left[\frac{m_p}{e}\frac{\bar{R}^{(0)}}{4(1+\bar{R}^{(0)})^2}
        \frac{1-\beta^3}{1+\beta^2}\right]
   \varepsilon^{i\mu\nu}D_\mu\Delta_b^{(I,1)}
   \frac{D_\nu\bar{\rho}^{(1)}_\gamma}{\bar{\rho}^{(0)}_\gamma}
\nonumber\\&&
- \left[\frac{m_p}{e}\frac{\bar{R}^{(0)}}{2(1+\bar{R}^{(0)})}
   \frac{1-\beta^3}{1+\beta^2}
   \frac{\dot{\bar{\rho}}^{(0)}_\gamma}{\bar{\rho}^{(0)}_\gamma}\right]
    \omega^{(2)i}.\label{App5}
\end{eqnarray}
We neglect the shear which is higher-order 
in cosmological perturbations.
First, we calculate the left-hand side. 
Using Eq.(\ref{App2}) and $\theta^{(0)}=3\cH/a$,
we find
\begin{eqnarray}
\dot{B}^{(2)i} + \frac{2}{3}\theta^{(0)} B^{(2)i}
&=&
  h^{i}_{~\mu} u_{(\gamma)}^\nu\nabla_\nu B^{(2)\mu}
+ 2\frac{\cH}{a}B^{(2)i}
\nonumber\\
&=&
  u_{(\gamma)}^{0}\nabla_0 B^{(2)i}
+ 2\frac{\cH}{a}B^{(2)i}
\nonumber\\
&=&\frac{1}{a}\left((B^{(2)i})'+3\cH B^{(2)i}\right)
\nonumber\\
&=&\frac{1}{a^4}\left(a^3B^{(2)i}\right)'
\label{App6}
.
\end{eqnarray}
Second, we calculate the second term in the right-hand side.
Using Eq.(\ref{ene0}) and Eq.(\ref{App2}), we obtain
\begin{eqnarray}
&&\left[\frac{m_p}{e}\frac{\bar{R}^{(0)}}{2(1+\bar{R}^{(0)})}
   \frac{1-\beta^3}{1+\beta^2}
   \frac{\dot{\bar{\rho}}^{(0)}_\gamma}{\bar{\rho}^{(0)}_\gamma}\right]
    \omega^{(2)i}
\nonumber\\
&&
\quad=
\left[\frac{m_p}{e}\frac{\bar{R}^{(0)}}{2(1+\bar{R}^{(0)})}
   \frac{1-\beta^3}{1+\beta^2}
   \frac{u_\gamma^0\dd_0\bar{\rho}^{(0)}_\gamma}
    {\bar{\rho}^{(0)}_\gamma}\right]
    \omega^{(2)i}
\nonumber\\&&\quad
=
- \left[\frac{m_p}{e}\frac{\bar{R}^{(0)}}{1+\bar{R}^{(0)}}
    \frac{1-\beta^3}{1+\beta^2}
  \right]
    \frac{2\cH}{a}\omega^{(2)i}.
\label{App7}
\end{eqnarray}
Third, we calculate the first term in the right-hand side.
Using Eq.(\ref{App4}) we get
\begin{eqnarray}
&&\left[\frac{m_p}{e}\frac{\bar{R}^{(0)}}{4(1+\bar{R}^{(0)})^2}
        \frac{1-\beta^3}{1+\beta^2}\right]
   \varepsilon^{i\mu\nu}D_\mu\Delta_b^{(I,1)}
   \frac{D_\nu\bar{\rho}^{(1)}_\gamma}{\bar{\rho}^{(0)}_\gamma}
\nonumber\\
&&=
- \left[\frac{m_p}{e}\frac{\bar{R}^{(0)}}{(1+\bar{R}^{(0)})^2}
         \frac{1-\beta^3}{1+\beta^2}\right]\times
\nonumber\\&&\qquad
    \times
    a^{-3}\epsilon^{ijk}\dd_j\Delta_b^{(I,1)}
    \left(\cH v_k^{(1)} -\frac{1}{4}
     \frac{\dd_k\delta\bar{\rho}_\gamma^{(1)}}{\bar{\rho}_\gamma^{(0)}}\right).
\label{App8}
\end{eqnarray}
Finally, we substitute Eqs.(\ref{App6}), (\ref{App7}), and (\ref{App8}) 
to Eq.(\ref{App5}), 
and then we get
\begin{eqnarray}
&&\left(a^3B^{(2)i}\right)'
\nonumber\\&&
=
\left[\frac{m_p}{e}\frac{\bar{R}^{(0)}}{(1+\bar{R}^{(0)})^2}
         \frac{1-\beta^3}{1+\beta^2}\right]\times
\nonumber\\&&\qquad
    \times
    a\epsilon^{ijk}\dd_j\Delta_b^{(I,1)}
    \left(\cH v_k^{(1)} -\frac{1}{4}
     \frac{\dd_k\delta\bar{\rho}_\gamma^{(1)}}{\bar{\rho}_\gamma^{(0)}}\right)
\nonumber\\&&\qquad
+ \left[\frac{m_p}{ae}\frac{1}{1+\bar{R}^{(0)}}
    \frac{1-\beta^3}{1+\beta^2}
  \right]
     \times2a^4\cH\omega^{(2)i}.
\label{App9}
\end{eqnarray}
Using Eq.(\ref{baralpha0}), Eq.(\ref{App9}) can be recast in
\begin{eqnarray}
&\left(a^3B^{(2)i}\right)'
=
\frac{1-\beta^3}{1+\beta}
    \frac{4\sigma_T}{3e}a^2\bar{\rho}_\gamma^{(0)}
\times\nonumber\\&\times
\left(\frac{2a^2\cH}{\bar{\alpha}^{(0)}}\omega^{(2)i}
+ \epsilon^{ijk}\frac{\bar{R}^{(0)}}{1+\bar{R}^{(0)}}\dd_j\Delta_b^{(I,1)}
    \delta v_{(\gamma b)k}^{(I,1)}
\right)
.
\end{eqnarray}
This result is the same as Eq.(\ref{MagneII}).



\end{document}